\documentclass[journal]{IEEEtran}
\usepackage{amsmath,amsfonts,amssymb,mathtools}
\usepackage[dvipsnames]{xcolor}

\ifCLASSINFOpdf
\else
   \usepackage[dvips]{graphicx}
\fi
\usepackage{url}

\usepackage{graphicx}
\usepackage{multirow}
\newcommand{\mycomment}[1]{}

\makeatletter
\def\ps@IEEEtitlepagestyle{
  \def\@oddfoot{\mycopyrightnotice}
  \def\@evenfoot{}
}
\def\mycopyrightnotice{
  {\footnotesize
  \begin{minipage}{\textwidth}
  \centering
  Copyright~\copyright~2023 IEEE. Personal use of this material is permitted. However, permission to use this  \\ 
  material for any other purposes must be obtained from the IEEE by sending a request to pubs-permissions@ieee.org.
  \end{minipage}
  }
}
\begin{document}

\title{Making Video Quality Assessment Models Robust to Bit Depth}

\author{Joshua P. Ebenezer, \IEEEmembership{Student Member, IEEE}, Zaixi Shang, \IEEEmembership{Student Member, IEEE}, Yongjun Wu, Hai Wei, Sriram Sethuraman and Alan C. Bovik, \IEEEmembership{Fellow, IEEE}
\thanks{Corresponding Author: Joshua Ebenezer (joshuaebenezer@utexas.edu). ORCID: 0000-0003-4936-9784.}
\thanks{This work was supported by a grant from Amazon and by grant number 2019844 for the National Science Foundation AI Institute for Foundations of Machine Learning (IFML). }
\thanks{J. P. Ebenezer, Z. Shang, and A. C. Bovik are with The Laboratory for Image and Video Engineering (LIVE) at The University of Texas at Austin, TX 78712 (e-mail: joshuaebenezer@utexas.edu).}
\thanks{Y. Wu, H. Wei, and S. Sethuraman are with Amazon Prime Video, Seattle, WA 98109.}}

\markboth{Signal Processing Letters}
{Shell \MakeLowercase{\textit{et al.}}: Bare Demo of IEEEtran.cls for IEEE Journals}
\maketitle

\begin{abstract}
We introduce a novel feature set, which we call HDRMAX features, that when included into Video Quality Assessment (VQA) algorithms designed for Standard Dynamic Range (SDR) videos, sensitizes them to distortions of High Dynamic Range (HDR) 
videos that are inadequately accounted for by these algorithms. While these features are not specific to HDR, and also augment the equality prediction performances of VQA models on SDR content, they are especially effective on HDR. HDRMAX features modify powerful priors drawn from Natural Video Statistics (NVS) models by enhancing their measurability where they visually impact the brightest and darkest local portions of videos, thereby capturing distortions that are often poorly accounted for by existing VQA models. As a demonstration of the efficacy of our approach, we show that, while current state-of-the-art VQA models perform poorly on 10-bit HDR databases, their performances are greatly improved by the inclusion of HDRMAX features when tested on HDR and 10-bit distorted videos.
\end{abstract}


\IEEEpeerreviewmaketitle

\section{Introduction}

\IEEEPARstart{H}{igh} Dynamic Range (HDR) videos have greater bit depths, wider color gamuts, and different opto-electronic transfer functions than Standard Dynamic Range (SDR) videos. For example, the HDR10 standard requires a bit depth of 10 bits, the BT 2020 color gamut (which covers 75.8\% of the perceiveable color gamut), and the Perceptual Quantizer (PQ) OETF, which is designed to represent luminances up to 10,000 nits. SDR videos are typically 8 bit, follow the BT 709 gamut, and use the gamma OETF, which is perceptually inaccurate for luminances above 100 nits. HDR videos also require displays that can radiate the increased ranges of brightnesses and colors, and specially designed hardware, whereas SDR videos were originally designed for Cathode-Ray Tubes, whose capabilities are exceeded by modern displays.  
\par Most state-of-the-art Video Quality Assessment (VQA) algorithms were designed to analyze SDR videos, primarily because the adoption and standardization of HDR is still ongoing. Existing VQA models are thus unable to capture distortions as they affect HDR over increased luminance/color ranges, and they perform poorly on predicting the quality of HDR videos. In a recent study~\cite{hdrchipqa}, we found that modifying the features of a powerful no-reference VQA model using an expansive nonlinearity significantly enhance its prediction performance on HDR video content. It has occurred to us that applying a similar concept to modify or augment other existing VQA models, including globally deployed algorithms, might also enhance their performances, which could significantly impact the delivery of high-quality HDR content worldwide. We therefore designed a new set of features, that we call HDRMAX, which can be used to augment both No-Reference (NR) VQA and Full-Reference (FR) models. We have found that including HDRMAX features improves robustness to bit depth in every case, dramatically improving their performances on distorted HDR video content. Remarkably, we have found that they also improve VQA model performance on distorted SDR videos although to a lesser degree, eliminating any concerns about their generalizability.

\section{Nonlinear features for VQA}

We will focus on processing transformed luminances (luma values) utilized in HDR standards like HDR10. Similar processing can be applied to Chroma values. In order to define HDRMAX features that are sensitized to distortions that affect the quality of HDR, begin by linearly scaling the luma values in each $W\times W$ spatial patch of each video frame to $[-1,1]$~\cite{hdrchipqa}. It may be observed that this scaling localizes the analysis of videos, and the resulting local dynamic ranges may significantly differ from the global. While this linear scaling to this particular range is not a requirement, it greatly simplifies the following processing steps and our exposition. Then, pass the linearly scaled luma values of each patch through an expansive nonlinearity 
\begin{equation}
\label{eq:exp_piecewise}
        f(x;\delta) =   \begin{cases} \exp(\delta x)-1 & x>0 \\
    1-\exp((-\delta x)) & x<0 \end{cases}.
\end{equation} 
\begin{figure}
    \centering
    \includegraphics[width=0.6\linewidth]{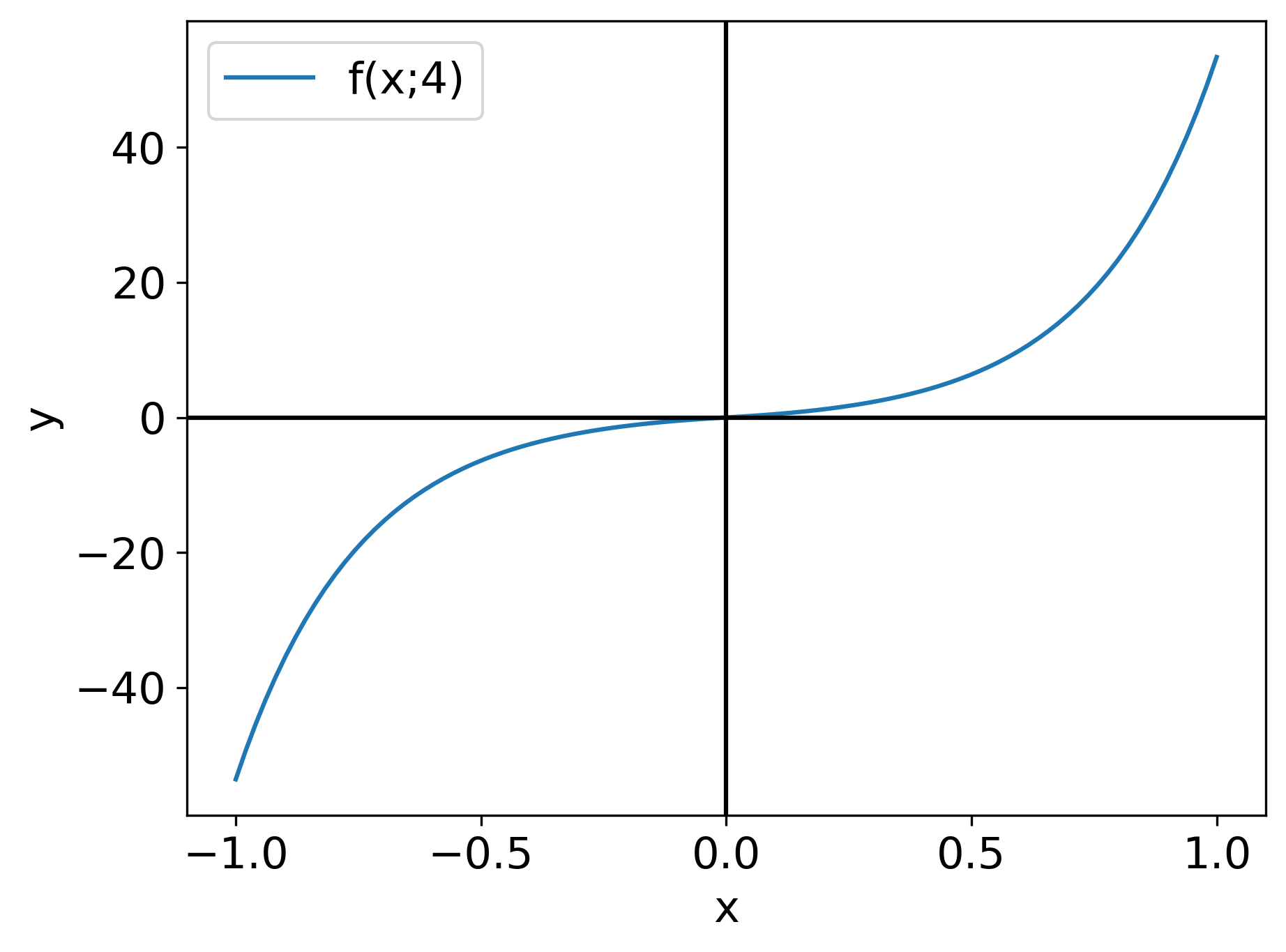}
    \caption{Plot of $f(x;4)$ vs $x$.}
    \label{fig:exp}
\vspace{-5mm}
\end{figure}
{The function is plotted in Fig.\ref{fig:exp}.}
This operation, which we refer to as the HDRMAX nonlinearity, expands the brightness ranges at the ends of the local dynamic luma scale while compressing the middle ranges. The rationale for (\ref{eq:exp_piecewise}) is simple but powerful. The features used in existing VQA algorithms are generally responsive to distortions that are endemic to videos, whether SDR or HDR. However, these responses are generally dominated by  distortions of the middle ranges of luma values, because they occupy most of the spatial extents of videos, as well as most of their dynamic ranges. Regions at the extremes of the dynamic range, which radiate very high or low (bright whites and dark blacks) luma (or chroma) values, distortions can be quite evident visually, but are often unaccounted for by standard VQA models because of dilution by the dominating responses to regions radiating brightnesses and colors in the middle portion of the dynamic range. The HDRMAX nonlinearity $f(x;\delta)$ counteracts this by isolating the extreme ends of the dynamic range at the expense of the middle range, which is suppressed. In this way, the modified feature responses are able to directly capture perceptual quality information more sensitive to bit-depth. As we shall see, including the HDRMAX nonlinearity to define new features to augment existing VQA models (while not replacing their original features) can produce dramatic improvements on distorted HDR videos, without sacrificing efficacy on SDR videos, with only a modest increment in complexity. 
 
\par The HDRMAX features are also defined using a second process, which we refer to as +NOISE, which involves applying a small amount of additive white Gaussian noise (AWGN) to the nonlinearly transformed luma values, thereby modeling the random signal fluctuations occurring at different stages along the visual pathway. {In addition, on regions that are smooth and lack texture, or are overexposed or underexposed, the addition of random noise provides numerical stability to processes that involve division by local energy or contrast estimates that underlie many VQA models.} Denoting a  frame processed by the HDRMAX nonlinearity as $P$, the resultant signal $Q$ (HDRMAX+NOISE) is given by
\begin{equation}
    Q[i,j] = P[i,j]+N(0,\sigma^2)
    \label{eq:noise}
\end{equation}
where $i,j$ are spatial pixel indices and $N(0,\sigma^2)$ is a zero-mean Gaussian random variable having standard deviation of $\sigma$. We fixed $\sigma=0.001$ in our experiments, although varying it in the range of $[5\times 10^{-4}, 5\times 10^{-3}]$ produces little changes in performance.

\subsection{HDRMAX Features for NR VQA}
 As already described, we modify NVS priors, using HDRMAX features to capture distortions on signal ranges visually expressive of HDR. Given the HDRMAX+NOISE processed luma values, extract quality-aware BRISQUE/NIQE (\cite{brisque,niqe}) features on them as follows. Model the distribution of the Mean-Subtracted Contrast Normalized (MSCN) coefficients~\cite{brisque} of $Q$ in (\ref{eq:noise}) as following a Generalized Gaussian Distribution (GGD). Using the maximum likelihood procedure deployed in BRISQUE/NIQE, find the shape parameter $\alpha$ and spread parameter $\sigma$ of the best GGD fit to the empirical distribution $Q[i,j]$. These serve as two features sensitized to video quality at the local extremes of dynamic range. 
 
\par Likewise, to capture the local correlation structure of $Q[i,j]$ (and the embedded distortions), also compute the products of neighboring pixels along four directions, then model the distributions of these products as following Asymmetric Generalized Gaussian Distributions (AGGD). In this way, an additional 16 features are arrived at ($\mu$,$\sigma_L$,$\sigma_R$, and $\gamma$, each along 4 orientations), computed exactly as they are in \cite{brisque,niqe}.

\mycomment{
as follows
\begin{align}\label{eq:pair}
\begin{split}
H[i,j,k_0] & = \hat{V}[i,j,k_0]\hat{V}[i,j+1,k_0] \\
V[i,j,k_0] & = \hat{V}[i,j,k_0]\hat{V}[i+1,j,k_0] \\
D_1[i,j,k_0] &= \hat{V}[i,j,k_0]\hat{V}[i+1,j+1,k_0] \\
D_2[i,j,k_0] & = \hat{V}[i,j,k_0]\hat{V}[i+1,j-1,k_0]
\end{split}
\end{align}

then model the distributions of these products as following Assymetric Generalized Gaussian Distributions (AGGD) of the form

\begin{equation}\label{eq:aggd}
g_2(x;\nu,\sigma_l^2,\sigma_r^2) = \begin{cases}
\frac{\nu}{(\beta_l+\beta_r)\Gamma (\frac{1}{\nu})} \exp(-(-\frac{x}{\beta_l})^\nu) &  x<0 \\
\frac{\nu}{(\beta_l+\beta_r)\Gamma (\frac{1}{\nu})} \exp(-(\frac{x}{\beta_r})^\nu) &  x>0,
\end{cases} 
\end{equation}
where
\begin{equation}
\beta_l = \sigma_l \sqrt{\frac{\Gamma(\frac{1}{\nu})}{\Gamma(\frac{3}{\nu})}} \quad \mathrm{and} \quad  \beta_r = \sigma_r \sqrt{\frac{\Gamma(\frac{1}{\nu})}{\Gamma(\frac{3}{\nu})}},
\end{equation}
and where $\nu$ controls the shape of the distribution and $\sigma_l$ and $\sigma_r$ control the spread on each side of the mode. The parameters ($\eta,\nu,\sigma_l^2,\sigma_r^2$) are extracted from the best AGGD fit to the histograms of each of the pairwise products in (\ref{eq:pair}), where 
\begin{equation}\label{eq:eta}
\eta = (\beta_r-\beta_l) \frac{\Gamma(\frac{2}{\nu})}{\Gamma(\frac{1}{\nu})}.
\end{equation}
}
The best fitting GGD and AGGD parameters are then averaged over all the frames of each video at two scales. In addition, the standard deviations of these parameters are computed over each non-overlapping group of five frames, then averaged over the entire video duration, yielding another feature set that is expressive of the  temporal variations of quality in the analyzed video. Thus, a total of 72 HDRMAX+NOISE features are found. These are used to supplement learning any given NR VQA model, as exemplified later.

\subsection{HDRMAX Features for FR VQA}

We derive a set of nonlinear HDRMAX features for FR VQA using the same nonlinear processing described in (\ref{eq:exp_piecewise}). Next, we extract the Visual Information Fidelity (VIF) features \cite{1576816} and the Detail Loss Metric (DLM) \cite{5765502} features directly on the HDRMAX processed luma values. We do not deploy the +NOISE modification, because VIF (and the related VQA engines ST-RRED \cite{soundararajan2012video} and SpEED-QA \cite{bampis2017speed}) already have neural noise models embedded in them. VIF is computed at the same four spatial scales as in VMAF \cite{vmaf}, and DLM is also calculated exactly as is done in \cite{vmaf}. These features are extracted from each frame, then averaged over all frames. No additional temporal features are found, since temporal effects are already modeled in the VIF features \cite{1576816}. 

\section{Databases}

We used two 10-bit databases and one 8-bit database to study the performance of HDRMAX-enhanced NR and FR VQA models. This allows us to guage the efficacy of HDRMAX on HDR content, while also ascertaining whether performance is maintained on SDR content.
\begin{itemize}
    \item The \textbf{LIVE HDR database~\cite{livehdr}}, a new database dedicated to the study of HDR perception, consisting of 310 videos viewed by 66 subjects.The distorted videos were created by compression using the x265 encoder and spatial scaling. All of the videos are compliant with the HDR10 standard.
    \item The \textbf{LIVE ETRI database}~\cite{etri} consists of 437 videos that had undergone compression, spatial aliasing, and temporal subsampling, along with human subjective scores. The videos in this database are 10 bit SDR, i.e., they are stored with a 10 bit representation but use the BT 709 gamma curve and color gamut.
    \item The \textbf{LIVE Livestream database}~\cite{livestream} contains 315 videos subjected to distortions commonly occurring in live streamed videos, such as interlacing, judder, compression, aliasing, frame-drops, flicker, and judder. Human scores are also supplied with this dataset. The videos in this database are 8 bit SDR.

\end{itemize}

\section{Results}
For each of the databases, we trained a Support Vector Regressor (SVR) against the given human opinion scores for each video. Each database was  divided into train and test sets using a 80:20 content separation. Cross-validation was performed to find the best parameters for the SVR on each database. The Spearman Rank Ordered Correlation Coefficient (SRCC) and Pearson's Linear Correlation Coefficient (PLCC) between the  predictions generated by each VQA model against the subjective quality scores were computed, and the median  values are reported over 100 random train-test splits.
Table~\ref{tab:results_hdr} gives the results obtained when HDRMAX+NOISE was included into existing NR VQA algorithms on the two 10 bit databases; the LIVE HDR dataset, and the LIVE ETRI 10 bit SDR database. As may be observed, enhancement with the HDRMAX+NOISE feature set improved performance of every algorithm on every database, often dramatically. VBLIINDS+HDRMAX+NOISE is the best performer on the LIVE HDR dataset, while ChipQA+HDRMAX+NOISE was the best performer on the ETRI dataset.

We also tabulated the results obtained by using the HDRMAX+NOISE feature set to enhance the performances of VQA algorithms on legacy 8 bit SDR video data in Table~\ref{tab:results_hdr}. Since HDRMAX+NOISE was designed to improve the quality prediction performance of existing VQA models when applied on HDR, we did not expect any significant improvement in performance. Instead, we were concerned that losses in performance could occur owing to feature dilution. Fortunately, in nearly every instance, the HDRMAX+NOISE-enhanced algorithms performed at least as well as they did without HDRMAX+NOISE augmentation on the LIVE Livestream database, and in most cases performed better. {Local contrast plays an important role in SDR video perception, as it does in HDR viewing, which helps explain HDRMAX's strong performance on SDR content as well.} These highly satisfactory results strongly suggest that HDRMAX+NOISE can be added to existing VQA algorithms to improve their robustness across bit depths. {On average, median SRCCs of NR VQA algorithms improved by 20\% on the LIVE HDR database, by 57\% on the ETRI 10-bit SDR database, and by 12\% on the LIVE Livestream 8-bit SDR database when the HDRMAX+NOISE features were included.}

\begin{table*}[!h]
\caption{Median SROCC and PLCC of NR Model performance against human judgments on 10 bit VQA Databases. Standard deviations are shown in parentheses. The best performing algorithm in each column is bold-faced.}
\begin{center} 
  \begin{tabular}{|p{4.5cm}|l|l|l|l|l|l|}
  \hline
   Dataset &
      \multicolumn{2}{c}{LIVE HDR} &
      \multicolumn{2}{c|}{LIVE ETRI (SDR 10 bit)}
      &  \multicolumn{2}{c|}{Livestream (SDR 8 bit)}\\
  \hline
 & SRCC & PLCC & SRCC & PLCC & SRCC & PLCC\\
   \hline
   HDRMAX & 0.8139\scriptsize(0.0584) & 0.7950\scriptsize(0.0534) & 0.6322\scriptsize(0.1242) & 0.6254\scriptsize(0.1367) 
 & 0.7658\scriptsize(0.0774) & 0.7735\scriptsize(0.0675)  \\
HDRMAX+NOISE & 0.7995\scriptsize(0.0643) & 0.7772\scriptsize(0.0941) & 0.6608\scriptsize(0.1042) & 0.6468\scriptsize(0.1076) &  0.7656\scriptsize(0.0854) & 0.7782\scriptsize(0.0722)   \\
\hline
TLVQM\cite{tlvqm} &   0.5781 \scriptsize(0.1014) & 0.5552 \scriptsize(0.0919) & 0.3398\scriptsize(0.1320) & 0.3184\scriptsize(0.1435)  &   0.7165\scriptsize(0.0868) & 0.7241\scriptsize(0.0800)  \\
TLVQM+HDRMAX & 0.8328\scriptsize(0.0591) & 0.8103\scriptsize(0.0570) & 0.7840\scriptsize(0.0759) & 0.6371\scriptsize(0.1318) &   0.8173\scriptsize(0.0736) & 0.8208\scriptsize(0.0692)  \\
TLVQM\cite{tlvqm}+HDRMAX+NOISE &  0.8194\scriptsize(0.0526) & 0.7992\scriptsize(0.0534)  &  0.6808\scriptsize(0.1265) & 0.6671\scriptsize(0.1255)&  0.8294\scriptsize(0.0745) & 0.8297\scriptsize(0.0687) \\
\hline
RAPIQUE\cite{rapique} & 0.4553 \scriptsize(0.2533) & 0.4864 \scriptsize(0.1171)&  0.1481\scriptsize(0.1419) & -0.0059\scriptsize(0.1633) & 0.3556\scriptsize(0.1637) & 0.3298\scriptsize(0.1545  \\
RAPIQUE\cite{rapique}+HDRMAX & 0.5458\scriptsize(0.1835) & 0.4363\scriptsize(0.3487)  & 0.4221\scriptsize(0.1745) & 0.3671\scriptsize(0.2749) & 0.3793\scriptsize(0.1550) & 0.3456\scriptsize(0.1469) \\
RAPIQUE\cite{rapique}+HDRMAX+NOISE & 0.6778\scriptsize(0.1583) & 0.6642\scriptsize(0.2645) & 0.4745\scriptsize(0.1619) & 0.4616\scriptsize(0.2971)  & 0.5083\scriptsize(0.1543) & 0.4643\scriptsize(0.2099) \\
\hline
BRISQUE\cite{brisque} & 0.7251 \scriptsize(0.0955) & 0.7139 \scriptsize(0.0881) & 0.3891\scriptsize(0.2109) & 0.3796\scriptsize(0.1966)  &  0.6564\scriptsize(0.1140) & 0.6840\scriptsize(0.1019)\\
BRISQUE\cite{brisque}+HDRMAX & 0.8024\scriptsize(0.0743) & 0.7763\scriptsize(0.0576) & 0.6323\scriptsize(0.1230) & 0.6308\scriptsize(0.1320) & 0.7535\scriptsize(0.0898) & 0.7728\scriptsize(0.0758) \\ 
BRISQUE\cite{brisque}+HDRMAX+NOISE & 0.8383\scriptsize(0.0737) & 0.8052\scriptsize(0.0816) & 0.6373\scriptsize(0.1136) & 0.6637\scriptsize(0.1348) &   0.7427\scriptsize(0.0771) & 0.7703\scriptsize(0.0664) \\
\hline
VIDEVAL\cite{videval} &  0.7131\scriptsize(0.1093) & 0.6513\scriptsize(0.1230)  & 0.6817\scriptsize(0.1181) & 0.6522\scriptsize(0.1199)&  0.8162\scriptsize(0.0679) & 0.8194\scriptsize(0.0588)  \\
VIDEVAL\cite{videval}+HDRMAX  &  0.8104\scriptsize(0.0682) & 0.7805\scriptsize(0.0672) & 0.6382\scriptsize(0.1318) & 0.6271\scriptsize(0.1368)& 0.8211\scriptsize(0.0657) & 0.8272\scriptsize(0.0602)  \\
VIDEVAL\cite{videval}+HDRMAX+NOISE & 0.8255\scriptsize(0.0632) & 0.7961\scriptsize(0.0604) &  0.6963\scriptsize(0.1161) & 0.6593\scriptsize(0.1150) &  0.8233\scriptsize(0.0725) & 0.8339\scriptsize(0.0639)  \\
\hline
VBLIINDS\cite{vbliinds} & 0.7484\scriptsize(0.1260) & 0.6930\scriptsize(0.1480) & 0.6121\scriptsize(0.1427) & 0.5729\scriptsize(0.1278)&  0.7464\scriptsize(0.0833) & 0.7572\scriptsize(0.0748)  \\
VBLIINDS\cite{vbliinds}+HDRMAX  & 0.8463\scriptsize(0.0748) & 0.8163\scriptsize(0.0700) & 0.6784\scriptsize(0.1247) & 0.6858\scriptsize(0.1250) &  0.8181\scriptsize(0.0600) & 0.8300\scriptsize(0.0527) \\
VBLIINDS\cite{vbliinds}+HDRMAX+NOISE & \textbf{0.8492\scriptsize(0.0751)} & \textbf{0.8190\scriptsize(0.1059)} & 0.7172\scriptsize(0.1032) & 0.7132\scriptsize(0.1061) &  0.8166\scriptsize(0.0670) & 0.8228\scriptsize(0.0590) \\
\hline 
HIGRADE\cite{higrade} &  0.6985\scriptsize(0.0839) & 0.6548\scriptsize(0.0866)  & 0.5018\scriptsize(0.1555) & 0.4675\scriptsize(0.1507)&  0.6801\scriptsize(0.0998) & 0.6913\scriptsize(0.0915)\\
HIGRADE+HDRMAX & 0.8074\scriptsize(0.0766) & 0.7881\scriptsize(0.0740) & 0.7927\scriptsize(0.0581) & 0.6632\scriptsize(0.1251) & 0.7654\scriptsize(0.0886) & 0.7681\scriptsize(0.0773)\\
HIGRADE\cite{higrade}+HDRMAX+NOISE & 0.8111\scriptsize(0.0690) & 0.7927\scriptsize(0.0674) & 0.6843\scriptsize(0.0963) & 0.6643\scriptsize(0.1108)& 0.7771\scriptsize(0.0894) & 0.7831\scriptsize(0.0773) \\
\hline
VSFA\cite{vsfa} &  0.7145\scriptsize(0.1070) & 0.6869\scriptsize(0.1000) & 0.6178\scriptsize(0.2050) & 0.6210\scriptsize(0.2303) &   0.8409\scriptsize(0.0684) & 0.8658\scriptsize(0.0561) 
\\
VSFA\cite{vsfa}+HDRMAX & 0.7414\scriptsize(0.0960) & 0.7196\scriptsize(0.0889) & 0.6014\scriptsize(0.2160) & 0.5789\scriptsize(0.2571) & 0.8439\scriptsize(0.0656) & 0.8652\scriptsize(0.0526)
\\
VSFA\cite{vsfa}+HDRMAX+NOISE     & 0.7518\scriptsize(0.0932) & 0.7315\scriptsize(0.0870) & 0.5955\scriptsize(0.2005) & 0.6111\scriptsize(0.2307) & \textbf{0.8472\scriptsize(0.0662)} & \textbf{0.8661\scriptsize(0.0519)}   \\
\hline
ChipQA\cite{chipqa} & 0.7435 \scriptsize(0.0895) & 0.7334 \scriptsize(0.0819) & 0.5934\scriptsize(0.1781) & 0.6033\scriptsize(0.1876)& 0.7980\scriptsize(0.0750) & 0.8066\scriptsize(0.0713) \\
ChipQA\cite{chipqa}+HDRMAX & 0.8117\scriptsize(0.0697) & 0.7766\scriptsize(0.0690)  & 0.7025\scriptsize(0.1370) & 0.7069\scriptsize(0.1685)& 0.8049\scriptsize(0.0729) & 0.8188\scriptsize(0.0643)  \\ 
ChipQA\cite{chipqa}+HDRMAX+NOISE & 0.8196\scriptsize(0.0698) & 0.7940\scriptsize(0.0693) & \textbf{0.7252\scriptsize(0.1278)} & \textbf{0.7275\scriptsize(0.1567)}&  0.8041\scriptsize(0.0670) & 0.8194\scriptsize(0.0661) \\
\hline
\end{tabular}
\label{tab:results_hdr}
\end{center}
\end{table*}

We also modified several leading FR algorithms {to include} HDRMAX, including peak signal-to-noise ratio (PSNR), SSIM, MS-SSIM \cite{wang2003multiscale}, VMAF \cite{vmaf}, SpEED-QA \cite{bampis2017speed}, and ST-RRED \cite{soundararajan2012video}. {We also included the HDR VQA model HDR-VDP-2 \cite{HDRVDP2} to show that the HDRMAX features are capable of improving the performance VQA models already designed to assess HDR content.} Since most FR VQA algorithms directly compute video quality predictions without using machine learning when mapping features to human opinion scores, we modified the compared FR VQA algorithms by breaking them into their constituent components, using each component as an individual feature. Specifically, we factored SSIM into three features: the three factors representing luminance, contrast and structural similarity features. Likewise, we decomposed MS-SSIM into eleven features, two SSIM features from each of four spatial scales, and three from the coarsest scale. On SpEED-QA, we extracted both the ``reduced-reference'' version and the ``single-number'' versions of the spatial and temporal SpEED-QA values to obtain four features. ST-RRED features were obtained from five levels of the Steerable Pyramid used in that algorithm. {HDR-VDP-2 features are the pooled quality features over nine spatial scales.} For each FR model, the combined feature set was obtained by concatenating the HDRMAX features with the original FR algorithm features. We then trained an SVR to map the features to human quality judgments on each of the 10-bit databases. As maybe observed in Table \ref{tab:results_FR_10}, every FR model modified by HDRMAX produced significantly improved quality prediction results on HDR videos, often by rather dramatic amounts. This includes SSIM, MS-SSIM, and VMAF models that are currently globally deployed by streaming and social media platforms. This suggests that HDRMAX can be used to improve the perceptual optimization of HDR video compression, potentially significantly reducing bandwidth consumption at the largest scales. To complete the analysis, we also studied the performances of the same models augmented by HDRMAX, but trained and tested on SDR videos. The results were computed on the LIVE Livestream database and are given in Table~\ref{tab:results_FR_10} as well. Again, no loss of performance was observed on the SDR video data; rather, there were improvements in all instances. {On average, the median SRCCs of FR VQA algorithms improved by 25\% on the LIVE HDR database, by 57\% on the ETRI 10-bit SDR database, and by 15\% on the LIVE Livestream 8-bit SDR database when the HDRMAX features were included.}

\begin{table*}[!h]
\caption{Median SROCC, LCC, and RMSE on 10 bit VQA Databases obtained using FR models. Standard deviations are shown in parentheses. The best performing algorithm is bold-faced.}
\begin{center} 
  \begin{tabular}{|p{4.5cm}|l|l|l|l|l|l|}
  \hline
   Dataset &
      \multicolumn{2}{c}{LIVE HDR} &
      \multicolumn{2}{c|}{LIVE ETRI (SDR 10 bit)}
      &  \multicolumn{2}{c|}{Livestream (SDR 8 bit)}\\
  \hline
 Algorithm & SRCC & PLCC & SRCC & PLCC & SRCC & PLCC\\
 \hline
HDRMAX          & {0.7681\scriptsize(0.0913)} & 0.7400\scriptsize(0.0958) & {0.8078\scriptsize(0.1035)} & 0.7950\scriptsize(0.0958) & {0.7535\scriptsize(0.0772)} & 0.7626\scriptsize(0.0702)  \\ 
PSNR            & \multicolumn{1}{l|}{0.6242\scriptsize(0.1504)} & 0.6357\scriptsize(0.1331) & {0.4941\scriptsize(0.1243)} & 0.4289\scriptsize(0.1289) & \multicolumn{1}{l|}{0.6063\scriptsize(0.0792)} & 0.6238\scriptsize(0.0716)  \\ 
PSNR+HDRMAX     & \multicolumn{1}{l|}{0.8263\scriptsize(0.0684)} & 0.8206\scriptsize(0.0615) & \multicolumn{1}{l|}{0.8268\scriptsize(0.1001)} & 0.8196\scriptsize(0.0940)  & \multicolumn{1}{l|}{0.7396\scriptsize(0.0432)} & 0.7488\scriptsize(0.0370)  \\ \hline
SSIM            & \multicolumn{1}{l|}{0.5208\scriptsize(0.1611)} & 0.4898\scriptsize(0.1595) 
 & \multicolumn{1}{l|}{0.3568\scriptsize(0.2625)} & 0.3358\scriptsize(0.2395) & \multicolumn{1}{l|}{0.6539\scriptsize(0.0927)} & 0.6584\scriptsize(0.0832)\\ 
SSIM+HDRMAX     & \multicolumn{1}{l|}{0.7771\scriptsize(0.0866)} & 0.7529\scriptsize(0.0964) & \multicolumn{1}{l|}{0.8485\scriptsize(0.0733)} & 0.8301\scriptsize(0.0741) & \multicolumn{1}{l|}{0.7521\scriptsize(0.0714)} & 0.7689\scriptsize(0.0619)  \\ \hline
MS-SSIM         & \multicolumn{1}{l|}{0.6007\scriptsize(0.1228)} & 0.5810\scriptsize(0.1260)  & \multicolumn{1}{l|}{0.5234\scriptsize(0.2336)} & 0.5319\scriptsize(0.2279) & \multicolumn{1}{l|}{0.7306\scriptsize(0.1097)} & 0.7377\scriptsize(0.1083)   \\ 
MS-SSIM+HDRMAX  & \multicolumn{1}{l|}{0.7645\scriptsize(0.0838)} & 0.7258\scriptsize(0.0868)  & \multicolumn{1}{l|}{0.7519\scriptsize(0.1399)} & 0.7297\scriptsize(0.1328) & \multicolumn{1}{l|}{0.7397\scriptsize(0.0712)} & 0.7724\scriptsize(0.0681)  \\ \hline
ST-RRED         & \multicolumn{1}{l|}{0.6863\scriptsize(0.0700)} & 0.6569\scriptsize(0.0744) & \multicolumn{1}{l|}{0.7500\scriptsize(0.0853)}   & 0.7587\scriptsize(0.0933) & \multicolumn{1}{l|}{0.6122\scriptsize(0.0738)} & 0.6273\scriptsize(0.0637)\\ 
ST-RRED+HDRMAX  & \multicolumn{1}{l|}{0.7896\scriptsize(0.0607)} & 0.7595\scriptsize(0.0603)& \multicolumn{1}{l|}{0.8628\scriptsize(0.0889)} & \textbf{0.8535\scriptsize(0.0840) } & \multicolumn{1}{l|}{\textbf{0.7685\scriptsize(0.0690)}}  & \textbf{0.7902\scriptsize(0.0630)}  \\ \hline
SpEED-QA        & \multicolumn{1}{l|}{0.611\scriptsize(0.1243)}  & 0.6196\scriptsize(0.1066) & \multicolumn{1}{l|}{0.7031\scriptsize(0.1485)} & 0.7179\scriptsize(0.1565) & \multicolumn{1}{l|}{0.5561\scriptsize(0.0481)} & 0.5891\scriptsize(0.0454) \\ 
SpEED-QA+HDRMAX & \multicolumn{1}{l|}{0.7581\scriptsize(0.0921)} & 0.7107\scriptsize(0.0993) &  \multicolumn{1}{l|}{0.8597\scriptsize(0.0971)} & 0.8355\scriptsize(0.0907) & \multicolumn{1}{l|}{0.6519\scriptsize(0.0416)} & 0.6642\scriptsize(0.0374)  \\ \hline
VMAF            & \multicolumn{1}{l|}{0.6753\scriptsize(0.0493)} & 0.6086\scriptsize(0.0583) & \multicolumn{1}{l|}{0.5617\scriptsize(0.0919)} & 0.5069\scriptsize(0.0844) & \multicolumn{1}{l|}{0.6424\scriptsize(0.0574)} & 0.7050\scriptsize(0.0498)  \\ 
VMAF+HDRMAX     & \multicolumn{1}{l|}{\textbf{0.8528\scriptsize(0.0543)}} & \textbf{0.8342\scriptsize(0.0632)} & \multicolumn{1}{l|}{\textbf{0.8654\scriptsize(0.1076)}}  &0.8417\scriptsize(0.0996) & \multicolumn{1}{l|}{0.7050\scriptsize(0.0853)} & 0.7120\scriptsize(0.0944) \\ \hline
HDR-VDP-2       & \multicolumn{1}{l|}{0.7041\scriptsize(0.1198)} & 0.6722\scriptsize(0.1081) & -& -& -& -\\ 
HDR-VDP-2+HDRMAX  & \multicolumn{1}{l|}{0.7431\scriptsize(0.0770)} & 0.7208\scriptsize(0.0764) & -& -& -& -\\ \hline
\end{tabular}
\label{tab:results_FR_10}
\end{center}
\end{table*}

\section{Conclusion}
The apparent success of our HDRMAX model shows that it is possible to build much stronger predictors of video quality for HDR, by employing it in existing SDR algorithms. Extensive evaluations on HDR and SDR validate these conclusions on both NR and FR VQA models. {Although the only currently available subjective HDR VQA databases utilize the HDR10 standard, HDRMAX is likely to be advantageous if applied to HDR standards like HDR10+ and Dolby Vision, which have an even greater capacity to represent local dynamic ranges than HDR10.}

\bibliographystyle{IEEEtran}
\bibliography{bare_jrnl}

\begin{thebibliography}{10}
\providecommand{\url}[1]{#1}
\csname url@samestyle\endcsname
\providecommand{\newblock}{\relax}
\providecommand{\bibinfo}[2]{#2}
\providecommand{\BIBentrySTDinterwordspacing}{\spaceskip=0pt\relax}
\providecommand{\BIBentryALTinterwordstretchfactor}{4}
\providecommand{\BIBentryALTinterwordspacing}{\spaceskip=\fontdimen2\font plus
\BIBentryALTinterwordstretchfactor\fontdimen3\font minus
  \fontdimen4\font\relax}
\providecommand{\BIBforeignlanguage}[2]{{%
\expandafter\ifx\csname l@#1\endcsname\relax
\typeout{** WARNING: IEEEtran.bst: No hyphenation pattern has been}%
\typeout{** loaded for the language `#1'. Using the pattern for}%
\typeout{** the default language instead.}%
\else
\language=\csname l@#1\endcsname
\fi
#2}}
\providecommand{\BIBdecl}{\relax}
\BIBdecl

\bibitem{hdrchipqa}
J.~P. Ebenezer, Z.~Shang, Y.~Wu, H.~Wei, S.~Sethuraman, and A.~C. Bovik,
  ``{HDR-ChipQA}: No-reference quality assessment on high dynamic range
  videos,'' \emph{Submitted to Elsevier Signal Process.: Image Comm.}

\bibitem{brisque}
A.~Mittal, A.~K. Moorthy, and A.~C. Bovik, ``No-reference image quality
  assessment in the spatial domain,'' \emph{IEEE Trans. Image Process.},
  vol.~21, no.~12, pp. 4695--4708, 2012.

\bibitem{niqe}
A.~Mittal, R.~Soundararajan, and A.~C. Bovik, ``Making a “completely blind”
  image quality analyzer,'' \emph{IEEE Signal Process. Lett.}, vol.~20, no.~3,
  pp. 209--212, 2012.

\bibitem{1576816}
H.~Sheikh and A.~Bovik, ``Image information and visual quality,'' \emph{IEEE
  Transactions on Image Processing}, vol.~15, no.~2, pp. 430--444, 2006.

\bibitem{5765502}
S.~Li, F.~Zhang, L.~Ma, and K.~N. Ngan, ``Image quality assessment by
  separately evaluating detail losses and additive impairments,'' \emph{IEEE
  Transactions on Multimedia}, vol.~13, no.~5, pp. 935--949, 2011.

\bibitem{soundararajan2012video}
R.~Soundararajan and A.~C. Bovik, ``Video quality assessment by reduced
  reference spatio-temporal entropic differencing,'' \emph{IEEE Transactions on
  Circuits and Systems for Video Technology}, vol.~23, no.~4, pp. 684--694,
  2012.

\bibitem{bampis2017speed}
C.~G. Bampis, P.~Gupta, R.~Soundararajan, and A.~C. Bovik, ``{SpEED-QA}:
  Spatial efficient entropic differencing for image and video quality,''
  \emph{IEEE Signal Process. Letters}, vol.~24, no.~9, pp. 1333--1337, 2017.

\bibitem{vmaf}
\BIBentryALTinterwordspacing
Netflix, \emph{VMAF: The Journey Continues}, 2018 (accessed January 13, 2023).
  [Online]. Available:
  \url{https://netflixtechblog.com/vmaf-the-journey-continues-44b51ee9ed12}
\BIBentrySTDinterwordspacing

\bibitem{livehdr}
Z.~Shang, J.~P. Ebenezer, A.~C. Bovik, Y.~Wu, H.~Wei, and S.~Sethuraman,
  ``Subjective assessment of high dynamic range videos under different ambient
  conditions,'' in \emph{IEEE Intl. Conf. Image Process.(ICIP)}, 2022, pp.
  786--790.

\bibitem{etri}
D.~Y. Lee, S.~Paul, C.~G. Bampis, H.~Ko, J.~Kim, S.~Y. Jeong, B.~Homan, and
  A.~C. Bovik, ``A subjective and objective study of space-time subsampled
  video quality,'' \emph{IEEE Trans. Image Process.}, vol.~31, pp. 934--948,
  2021.

\bibitem{livestream}
Z.~Shang, J.~P. Ebenezer, Y.~Wu, H.~Wei, S.~Sethuraman, and A.~C. Bovik,
  ``Study of the subjective and objective quality of high motion live streaming
  videos,'' \emph{IEEE Trans. Image Process.}, vol.~31, pp. 1027--1041, 2022.

\bibitem{tlvqm}
J.~Korhonen, ``Two-level approach for no-reference consumer video quality
  assessment,'' \emph{IEEE Trans. Image Process.}, vol.~28, no.~12, pp.
  5923--5938, 2019.

\bibitem{rapique}
Z.~Tu, X.~Yu, Y.~Wang, N.~Birkbeck, B.~Adsumilli, and A.~C. Bovik, ``{RAPIQUE}:
  Rapid and accurate video quality prediction of user generated content,''
  \emph{IEEE Open J. Signal Process.}, vol.~2, pp. 425--440, 2021.

\bibitem{videval}
Z.~Tu, Y.~Wang, N.~Birkbeck, B.~Adsumilli, and A.~C. Bovik, ``{UGC-VQA}:
  Benchmarking blind video quality assessment for user generated content,''
  \emph{IEEE Trans. Image Process.}, vol.~30, pp. 4449--4464, 2021.

\bibitem{vbliinds}
M.~A. Saad, A.~C. Bovik, and C.~Charrier, ``Blind prediction of natural video
  quality,'' \emph{IEEE Trans. Image Process.}, vol.~23, no.~3, pp. 1352--1365,
  2014.

\bibitem{higrade}
D.~Kundu, D.~Ghadiyaram, A.~C. Bovik, and B.~L. Evans, ``No-reference quality
  assessment of tone-mapped {HDR} pictures,'' \emph{IEEE Trans. Image
  Process.}, vol.~26, no.~6, pp. 2957--2971, 2017.

\bibitem{vsfa}
D.~Li, T.~Jiang, and M.~Jiang, ``Quality assessment of in-the-wild videos,'' in
  \emph{Proceedings of the 27th ACM International Conference on Multimedia},
  2019, pp. 2351--2359.

\bibitem{chipqa}
J.~P. Ebenezer, Z.~Shang, Y.~Wu, H.~Wei, S.~Sethuraman, and A.~C. Bovik,
  ``{ChipQA}: No-reference video quality prediction via space-time chips,''
  \emph{IEEE Trans. Image Process.}, vol.~30, pp. 8059--8074, 2021.

\bibitem{wang2003multiscale}
Z.~Wang, E.~P. Simoncelli, and A.~C. Bovik, ``Multiscale structural similarity
  for image quality assessment,'' in \emph{The Thrity-Seventh Asilomar
  Conference on Signals, Systems \& Computers, 2003}, vol.~2.\hskip 1em plus
  0.5em minus 0.4em\relax Ieee, 2003, pp. 1398--1402.

\bibitem{HDRVDP2}
R.~Mantiuk, K.~J. Kim, A.~G. Rempel, and W.~Heidrich, ``{HDR-VDP-2}: A
  calibrated visual metric for visibility and quality predictions in all
  luminance conditions,'' \emph{ACM Trans. Graphics (TOG)}, vol.~30, no.~4, pp.
  1--14, 2011.

\end{thebibliography}

\end{document}